\documentclass{article}
\usepackage{graphicx}
\usepackage{amsmath}
\usepackage{amsfonts}
\usepackage{amssymb}

\begin{document}

\begin{center}
{\Large A quantum algorithm providing exponential speed increase for finding
eigenvalues and eigenvectors }\footnote{This work supported in part by grant
\# N00014-95-1-0975 from the Office of Naval Research; by ARO and DARPA under
grant \# DAAH04-96-1-0386 to QUIC, the Quantum Information and Computation
initiative; by a DARPA grant to NMRQC, the Nuclear Magnetic Resonance Quantum
Computing initiative; and by ARO through an NDSEG fellowship.}

\bigskip\medskip

Daniel S. Abrams

Department of Physics, MIT 12-128b

Cambridge, MA 02139 (abrams@mit.edu)

\medskip\medskip

Seth Lloyd

d'Arbeloff Laboratory for Information Sciences and Technology

Department of Mechanical Engineering, MIT 3-160

Cambridge, MA 02139 (slloyd@mit.edu)

\medskip\medskip
\end{center}

We describe a new polynomial time quantum algorithm that uses the quantum fast
fourier transform to find eigenvalues and eigenvectors of a Hamiltonian
operator, and that can be applied in cases (commonly found in ab initio
physics and chemistry problems) for which all known classical algorithms
require exponential time. Applications of the algorithm to specific problems
are considered, and we find that classically intractable and interesting
problems from atomic physics may be solved with between 50 and 100 quantum bits.

\bigskip

\bigskip

\bigskip

\bigskip

\bigskip

\bigskip

\bigskip

\bigskip

\bigskip

\bigskip

\bigskip

\bigskip

\bigskip

\bigskip\newpage

Long before Shor's ground-breaking algorithm\cite{SHOR} - and the resulting
surge of interest in quantum computing - Feynman suggested that a quantum
computer might be useful for simulating other quantum systems\cite{Feynman}.
This suggestion was based upon the observation that quantum systems are
described in a Hilbert space whose size grows exponentially with the number of
particles. Thus a collection of only 100 spin 1/2 particles, each of which
could be specified by only two complex amplitudes were it isolated,\ requires
a total of 2$^{100}$ complex amplitudes for its state to be specified
completely. This exponential explosion severely limits our ability to perform
true ``ab initio''\ (first principles) calculations; since it is obviously not
possible to even describe the state of anything but the smallest quantum
systems, one must resort to various approximation techniques to calculate
properties of interest.\ 

Recent work in quantum computation has revealed various techniques for
$simulating$ physics on a quantum computer \cite{Lloyd Sim}\cite{Abrams}%
\cite{Boghosian}\cite{Zalka}\cite{Wiesner}\cite{Lidar}, and it has been
demonstrated that this can in fact be accomplished efficiently, as Feynman
supposed. However, while previous work has described a variety of algorithms
for initializing a quantum computer into a state corresponding to the state of
a physical system, for time evolving this state on the computer, and for
measuring properties of the time-evolved state \cite{Lloyd Sim}\cite{Abrams}
\cite{Boghosian}\cite{Zalka}\cite{Wiesner}, there has been comparatively
little work done on algorithms which $calculate$ $static$ $properties$ of a
physical system\cite{Lidar}. In particular, of all the questions which one
might ask about a quantum system, there is one most frequently asked and for
which one would most greatly desire an efficient algorithm: What are the
energy eigenvalues and eigenstates? In this letter, we provide a quantum
algorithm that can find eigenvalues and eigenvectors of a Hamiltonian operator
in cases that occur frequently in problems of physical interest. Moreover, the
algorithm requires an amount of time which scales as a polynomial function of
the number of particles and the desired accuracy, whereas all known classical
algorithms require an exponential amount of time.\ 

The problem to be solved can be precisely stated as follows.\ Consider the
time-evolution operator $\widehat{U}$ = $e^{-\frac{i}{\hslash}\widehat{H}t}%
$\ which corresponds to the Hamiltonian $\widehat{H}$, and an approximate
eigenvector V$_{a}$ of $\ \widehat{U}$ \ (and thus of $\widehat{H}$)\ that can
be generated in quantum polynomial time, i.e., the machine can be placed into
a state corresponding to $V_{a}$ using a polynomial number of quantum logic
operations. Call the true eigenvector $V$ and the true eigenvalue $\lambda
_{v}.$ If the state $V_{a}$ satisfies the property that $\left|  \left\langle
V_{a}|V\right\rangle \right|  ^{2}$ is not exponentially small - that is, the
approximate eigenvector contains a component of the actual eigenvector that is
bounded by a polynomial function of the problem size - then $V$ and
$\lambda_{v}$ can be found in time proportional to $1/\left|  \left\langle
V_{a}|V\right\rangle \right|  ^{2}$ and $1/\epsilon$, where $\epsilon$ is the
desired accuracy.

Intuitively, what the algorithm does is to resolve the guess into its
non-negligible components and determine the corresponding eigenvalues. If the
operator $\widehat{U}$ (and thus its eigenvectors) is of exponentially large
dimension - which it typically is - there are no known classical algorithms
that can find even the eigenvalues in polynomial time. Although the
requirement that there exist an initial statevector $V_{a}$ with the specified
properties may appear to be overly restrictive, it is frequently (if not
usually) possible to obtain such a guess for ``real'' problems using existing
classical techniques. For example, in any physical system with discrete energy
levels that are not exponentially close together near the ground state (such
as an atom), if it is possible to obtain classically any state vector with
expected energy merely less than the first excited state (by a
non-exponentially small amount), then this state vector must contain a
non-negligible component of the ground state and - although it may not
remotely resemble the ground state - could be used as the approximate state
V$_{a}$ to determine the true ground state and ground state energy in
polynomial time. Finally, if for some problems it is not possible to obtain
classically a guess with the desired properties, it may often be the case that
the state vector $V_{a}$ may be generated using a quantum algorithm, such as
quantum simulated annealing.

We will now describe an algorithm which applies to any $\widehat{U}$ that can
be implemented in quantum polynomial time, whether or not it represents the
time evolution operator corresponding to a given Hamiltonian. (It was shown in
\cite{Lloyd Sim} that the time evolution operator corresponding to any local
Hamiltonian can be implemented in polynomial time on a quantum computer.) This
first part of the algorithm was described independently by Cleve et. al. in
\cite{Cleve} to find eigenvalues (but not eigenvectors) of unitary operators,
in that case because the eigenvalues of a particular operator can be used to
solve the abelian stabilizer problem. To begin, we consider a quantum computer
consisting of m+l+w qubits, where a total of m qubits (to be called the index
bits)\ are used for an FFT, a total of l qubits describe the Hilbert space in
which the operator $\widehat{U}$ acts, and w extra working qubits are required
for temporary storage. Let M = $2^{m}$. The accuracy of the result will grow
as 1/M. \ Assume that the m index qubits are initially in the state
$\vert$%
0%
$>$%
and that the l qubits are initially in the state V$_{a}$ (hence, the need for
V$_{a}$ to be generated in quantum polynomial time). That is, the initial
state is\bigskip%
\begin{equation}
|\Psi>=|0>\ |V_{a}>
\end{equation}

where the w work qubits are assumed to be
$\vert$%
0%
$>$%
unless specified otherwise. We perform a $\pi/2$ rotation on each of the m
index qubits to obtain the state%

\begin{equation}
|\Psi>=\frac{1}{\sqrt{M}}\sum\limits_{j=0}^{M-1}|j\rangle|V_{a}\rangle
\end{equation}

\bigskip Next, one performs a series of quantum logic operations that
transform the computer into the state%

\begin{equation}
|\Psi>=\frac{1}{\sqrt{M}}\sum\limits_{j=0}^{M-1}|j\rangle(\widehat{U}%
)^{j}|V_{a}\rangle\label{Goal}%
\end{equation}
This transformation is accomplished by applying the operation $\widehat{U}$ to
the second set of l qubits (which are initially in the state $|V_{a}\rangle$)
j times. It can be implemented easily by performing a loop (indexed by i) from
1 to $M$. Using standard quantum logic operations, set a flag qubit to the
value
$\vert$%
1%
$>$%
if and only if \ i
$<$%
j and perform the operation $\widehat{U}$ conditioned on the value of this
flag. Thus only those components of the above superposition for which i%
$<$%
j are effected. Finally, undo the flag qubit and continue with the next
iteration. \bigskip After M iterations, the state above is obtained.

At this point, it is helpful to rewrite the state in a slightly different
manner. Label the eigenvectors of $\widehat{U}$ by the states $\left|
\phi_{k}\right\rangle $ and the corresponding eigenvalues with $\lambda_{k}$.
We can then write%

\begin{equation}
|V_{a}\rangle=\sum\limits_{k}^{{}}c_{k}\left|  \phi_{k}\right\rangle
\end{equation}

in which case the state (\ref{Goal}) above can be rewritten as\bigskip%
\begin{align}
|\Psi\rangle &  =\frac{1}{\sqrt{M}}\sum\limits_{j=0}^{M-1}|j\rangle
(\widehat{U})^{j}\sum\limits_{k}^{{}}c_{k}\left|  \phi_{k}\right\rangle \\
&  =\frac{1}{\sqrt{M}}\sum\limits_{k}^{{}}c_{k}\sum\limits_{j=0}%
^{M-1}|j\rangle(\lambda_{k})^{j}\left|  \phi_{k}\right\rangle
\end{align}

If we write $\lambda_{k}$ as$\ e^{i\omega_{k}}$ and exchange the order of the
qubits so that the labels $\left|  \phi_{k}\right\rangle $ appear first, the
result is seen then most clearly:\bigskip%
\begin{equation}
|\Psi\rangle=\frac{1}{\sqrt{M}}\sum\limits_{k}^{{}}c_{k}\left|  \phi
_{k}\right\rangle \sum\limits_{j=0}^{M-1}e^{i\omega_{k}j}|j\rangle
\end{equation}

It is now self-evident that a quantum FFT performed on the m index qubits will
reveal the phases $\omega_{k}$ and thereby the eigenvalues $\lambda_{k}$. The
quantum FFT requires only poly(m) operations, whereas the accuracy of the
result will scale linearly with M or 2$^{m}$. Each frequency is seen to occur
with amplitude $c_{k}$ $=$ $\left\langle V_{a}|\phi_{k}\right\rangle $; by
performing a measurement on the m index qubits, one thus obtains each
eigenvalue with probability $\left|  c_{k}\right|  ^{2}$. \ Only a polynomial
number of trials is therefore required to obtain any eigenvalue for which
$c_{k}$ is not exponentially small. If the initial guess $|V_{a}\rangle$ is
close to the desired state (i.e., $\left|  <V_{a}|V>\right|  ^{2}$ is close to
1), then only a few trials may be necessary.

Moreover, one obtains the eigenvectors as well:\ once a measurement is made
and an eigenvalue $\lambda_{k}$ is determined, the remaining l qubits
``collapse'' into the state of the corresponding eigenvector. Of course, the
state $\left|  \phi_{k}\right\rangle $ is in some sense ``trapped'' inside the
computer. But since it is impossible to store as classical information the
2$^{l}$ phases associated with the state, one cannot possibly hope to do
better. However, one is likely to be interested in various properties of the
eigenvectors, and these can be determined by making various measurements on
the state. For ab initio quantum calculations, easily obtainable properties
include those of greatest interest: charge density distributions, correlation
functions, momentum distributions, etc. See \cite{Abrams} for a discussion of
how relevant physical information can be extracted efficiently from the
quantum computer.

We now consider more precisely how to find the eigenvectors and eigenvalues of
a ``real''\ Hamiltonian. Generally, one wishes to find energy eigenstates for
a Hamiltonian of the form%

\begin{equation}
H=\sum\limits_{i=1}^{n}(T_{i}+V_{i})+\sum\limits_{i>j}^{n}V_{ij}%
\end{equation}

where $n$ is the number of particles, $T_{i}$ is the kinetic energy, $V_{i}$
is the external potential, and $V_{ij}$ is the interaction between the
particles. However, there is no reason why these techniques cannot be applied
to a different Hamiltonian or to one containing additional terms, as long as
the Hamiltonian can be separated into a sum of local interactions (that is, a
sum of terms which act upon only k qubits, where k is independent of the
number of particles n). (In atomic problems, for example, one might include
effective interactions such as spin-orbit coupling or nuclear finite size
effects). \ Because the Hamiltonian is Hermitian, we apply the steps described
above to the time evolution operator $\widehat{U}(t)$ = $e^{-iHt}$, which is
unitary and has the same eigenvalues and eigenvectors. This time evolution
operator is generated using the technique described in \cite{Lloyd Sim}; the
key idea is to write%

\begin{align}
H &  =\sum\limits_{{}}^{{}}H_{i}\\
\widehat{U}(t) &  =e^{-iHt}=(e^{-iH_{1}\frac{t}{m}}e^{-iH_{2}\frac{t}{m}%
}...e^{-iH_{k}\frac{t}{m}})^{m}+\sum\limits_{i>j}^{{}}[H_{i},H_{j}]\frac
{t^{2}}{2m}+...
\end{align}

where each $H_{i}$ acts on only k qubits at a time. (In the Hamiltonian above,
each $H_{i}$ represents one of the terms $T_{i}$, $V_{i}$, or $V_{ij})$. Let
$U_{i}=$ $e^{-iH_{i}\frac{t}{m}}$. Each term $U_{i}$ can be implemented
efficiently, because it acts in a space of only k quantum bits, where k is
small. For large enough $m$, the second term on the right (and the higher
order terms) approaches zero. It is therefore possible to generate
$\widehat{U}(t)$ by acting on the state with each $U_{i}$ in series, a total
of $m$ times. A careful analysis \cite{Lloyd Sim} reveals that in order to
simulate $\widehat{U}(t)$ with an accuracy $\epsilon$, one needs to apply
O($t^{2}/\epsilon$) quantum logic operations.\footnote{Since U(t) has the same
eigenvalues and vectors for all t, this might lead one to falsely conclude
that the number of operations necessary to find the eigenstates to a given
accuracy could be reduced by choosing a shorter length of time t for the
operator U(t). However, the algorithm requires one to calculate U$^{M}$, and
since U(t)$^{M}$ = U(Mt), one sees that U = U(t) must be calculated with
greater precision if U$^{M}$ is to be calculated for a fixed precision. In
fact, since the eigenvectors are determined with a precision proportional to
M, the number of quantum logic operations required to calculate the energy
eigenstates to a precision $\epsilon$ is seen to scale as $\epsilon^{-2}$.}

For a specific problem, the form of the matrices $U_{i}$ depends greatly on
the basis set chosen to describe the Hilbert space. Moreover, the choice may
strongly impact the size of the basis required to describe the system
accurately. In the usual first quantized representation, each particle is
described by a series of l qubits representing a single particle wave
function. The system as a whole is thus represented with n*l qubits. (It is
also possible to use a second quantized representation, which may be more
efficient for certain problems; see \cite{Abrams}). For the Hamiltonian above,
the matrices $U_{i}$ can be implemented in a particularly efficient manner by
using either position space or momentum space for the single particle basis,
and switching between the two via quantum FFTs. However, for most problems,
these are not the most effective choices to represent the energy eigenstates.
Other sets of basis states are generally more efficient and are frequently
employed in conventional classical computations: one possible example might be
wavelets; another common choice might be single electron solutions for an
effective potential. As long as the single particle basis is of a fixed size
(and the reason why we choose a more complicated basis set is for the explicit
purpose of keeping it small), then the operators $U_{i}$ can always be
calculated in the chosen basis and implemented using O(d$^{4}$) operations,
where d is the dimension of the $single$ $particle$ basis set \cite{Barenco}.
Thus one finds that it is possible to apply these quantum algorithms using the
more elaborate choices of basis sets that are commonly employed in
conventional ab initio calculations.(Because there exists a fast quantum
wavelet transform \cite{Williams}, it may be that a wavelet basis turns out to
be particularly useful). On the other hand, there is a trade-off between
memory and speed. By using the position or momentum space representation, one
needs only O(poly(k)) = O(poly(log d)) operations to perform each $U_{i}$;
however a large number of qubits are required to describe the eigenstates
accurately. By choosing a more elaborate basis set, one can vastly reduce the
required number of qubits, but a much larger number of quantum logic
operations O(k$^{4}$) may be necessary to implement each $U_{i}$. Thus one
finds that, just as with conventional computations, the choice of basis sets
in the quantum computation will depend upon the specific problem at hand and
the specific capabilities of the actual computing machine.

Normally, the initial state $V_{a}$ will be the result of a classical
calculation, for example, a Hartree-Fock calculation or configuration
interaction calculation. Any ab initio technique which results in a known wave
function can be used. (Note that this does not include those techniques which
utilize density functional theory, as we require a wavefunction, not simply a
charge density distribution). If the input wave function is not already
symmetrized or antisymmetrized, we can use the algorithms described in
\cite{Abrams} to do so efficiently.

Finally, we consider state-of-the-art ab initio calculations of atomic energy
levels in order to compare the quantum algorithm described above with known
classical techniques. Problems from atomic physics serve as a particularly
good benchmark because extremely accurate experimental data is widely
available. The quantum algorithm corresponds most closely to what is known as
``complete active configuration interaction'' or ``full configuration
interaction'' techniques, because the many-particle basis set includes all
possible products of single particle basis vectors. \ This approach is most
valuable in situations where the correlation energy is large and where many
``configurations'' are of similar energy (this typically occurs when many
electrons are in open shells). Unfortunately, it is difficult to state
precisely the minimum size problem for which the quantum calculation surpasses
the best classical calculations, because a variety of sophisticated techniques
are used to avoid the exponential explosion in basis states. That is, the most
$accurate$ classical calculations do not employ directly the ``full
configuration interaction'' method. Based on \cite{Johnson}, however, we
estimate that a calculation of the energy levels of B (5 electrons), using
roughly 20 angular wavefunctions and 40 radial wavefunctions per particle -
for a\ total of 800 single particle wave functions and therefore 800$^{5}$
$\thickapprox$10$^{15}$ full many-body basis states - may provide more
accurate results than any classical calculation performed to date. At the very
least, such a calculation would reveal scientifically interesting (and
classically unobtainable) results with respect to electron correlation
energies in B and the relative importance of various orders of excited configurations.

A quantum calculation of the B ground state, using a basis set as described
above, can be accomplished with 60 qubits:\ 10 per particle to represent the
state of the atom (for a total of 50 qubits), 6 or 7 qubits for the FFT, and a
few additional ``scratch'' qubits\footnote{The number of qubits required for
the FFT is not as large as one might at first suppose, based on the earlier
statement that the accuracy scales linearly with the size of the FFT. This
statement is true only for a fixed U. By changing U - in particular, by
increasing the length of time t in U(t) - one can obtain the eigenvalues to
arbitrary precision using a fixed number of FFT points. However, the number of
points in the FFT must be sufficiently large so as to seperate the frequencies
corresponding to distinct eigenvectors. This is how the estimate of 6 or 7
qubits\ (64 or 128 FFT points) is made.}. Unfortunately, the two particle
operators (generated by the coulomb attraction between pairs of electrons)
take place in a subspace of dimension $(2^{10})^{2}$; they therefore are
represented by matrices with $2^{40}$ elements. Implementing such an operator
by brute force is likely to remain intractable for the foreseeable future.
However, it may be possible to perform the necessary transformation using a
quantum algorithm. One possible technique is to change basis sets: by
representing the interacting particles in position space, instead of with the
orbital basis set, it is easy to calculate the coulomb terms. Thus one can
transform each particle into position space separately (requiring a small
number of quantum logic operations), perform the time evolution corresponding
to the coulomb interaction, and then transform back. Unfortunately, a position
space representation will require many more qubits. We estimate that 30 qubits
per particle (10 per dimension, for a real space grid of 1024x1024x1024 per
particle) \ will more than suffice. Because these 30 qubits are required only
temporarily for the 2 particles whose interaction we are considering at any
particular stage in the algorithm, the new efficient algorithm requires a
total of 2 x 30 qubits (for the interacting particles), an additional 3 x 10
qubits (for the remaining particles), and the same 10 qubits for the FFT and
work space. It thus appears that in order to realistically perform an
``interesting'' calculation using the algorithms described previously, one
will need a quantum computer with approximately 100 qubits. Of course, the
possibility remains that an efficient algorithm for implementing the coulomb
interaction could be invented that does not require additional working space.

In conclusion: we have provided a new quantum algorithm which can be used to
find eigenvectors and eigenvalues of a Hamiltonian operator. The algorithm
provides an exponential speed increase when compared to the best known
classical techniques. Problems from atomic physics may be the best place to
perform the first real calculations, both because accurate experimental data
is available to verify the resulting calculations, and because the parameters
involved appear to be within the foreseeable range of small quantum computers.
We estimate that 50 - 100 qubits are sufficient to perform ``interesting''
calculations that are classically intractable. \ Finally, we suggest a couple
of interesting questions which remain open. First; although we have made
estimates regarding numbers of required qubits, it would be interesting to
calculate accurately the number of quantum logic gates required to do an
``interesting'' problem. Second, a detailed analysis of the effects of errors
would be worthwhile, as would an analysis of error correcting codes in this context.

D.S.A. acknowledges support from a NDSEG fellowship, and thanks D. Lidar, C.
Froese Fisher, and especially W. R. Johnson for helpful discussions. Portions
of this research were supported by grant \# N00014-95-1-0975 from the Office
of Naval Research, and by ARO and DARPA under grant \# DAAH04-96-1-0386 to
QUIC, the Quantum Information and Computation initiative, and by a DARPA grant
to NMRQC, the Nuclear Magnetic Resonance Quantum Computing initiative.

\end{document}